# Product Cycle, Wintelism, and Cross-national Production Networks (CPN) for Developing Countries

## -- China's Telecom Manufacturing Industry as A Case.

**Zixiang (Alex) Tan, Ph.D.**
**Syracuse University**
ztan@syr.edu
**September, 2001**

# 1. Introduction

China has recently emerged as a significant player in the global telecommunications industry. Its enormous market size has attracted nearly all the significant manufacturers in the world to engage in various trade and production activities. Direct imports from advanced countries continue to support most of China's high-end market. Local subsidiaries and joint ventures of multinational telecom manufacturers have grown to supply a large percentage of the medium-end of Chinese market. Meanwhile, indigenous producers have recently emerged to dominate the low-end market and to aggressively compete in the medium-end market. The coexistence of these three types of suppliers results from a dynamic and complex interaction among multinational corporations, governments, and China's indigenous industry. Understanding the interaction has long been a research topic for academic investigators as well as a practical question for decision-makers in investing and hosting governments, multinational corporations, and local manufacturers. On the one hand, the interaction is currently shaping the competition in today's telecom market in China. On the other hand, the emerging manufacturing base in China positions it to have an inevitable influence on the global telecom manufacturing industry. The impacts on the global market will be further accelerated when China join the WTO in later 2001.

Much research has been done regarding the interaction among multinational corporations, governments, and indigenous industries in Asian countries, especially in the electronics industry. Product cycle theory, Wintelism, and cross-national production networks among others have been developed to examine and explain the interaction and industrialization in Asian countries and their impacts on global competition. The original product cycle theory was developed by Japanese economist Akamatsu Kaname (1937, in Japanese). Several scholars later expanded product cycle theory (Vernon, 1971; Cumings, 1984; Kojima, 1986). Today's product cycle theory argues that new products are first manufactured in advanced countries and the demand in other nations is met through exports. Production is later moved to other nations to take advantage of cheap labor costs and/or to break market entry barriers. Finally, the global demand is met by producers in countries where the production costs are lower and the original producers in advanced countries often abandon their production (Bernard & Ravenhill, 1995). Product cycle theory predicts a gradual industrialization and a base establishment for export in less developed countries.

Wintelism, derived from Windows and Intel, reflects the success of Microsoft's Windows in operating systems and Intel's CPU in the computer processor market. Wintelism argues that market competition has shifted away from final assembly and vertical control of markets by traditional corporations such as IBM as well as Japanese Keiretsu trading groups. Taking the computer industry as an example, Wintelism sees that current market competition can happen anywhere in the production value-chain, including components, subsystems, system assembly, operation software, and applications software. These traditionally integrated system elements "become separate and critical competitive markets" (Borrus & Zysman, 1997). Companies like Intel in components, Compaq in PCs, Microsoft in operating systems, and Adobe in applications software could all compete and flourish in their specialized and separate markets within the computer industry.

Cross-national Production Networks (CPN) refer to "the consequent dis-integration of the industry's value chain into constituent functions that can be contracted out to independent producers wherever those companies are located in the global economy" (Borrus & Zysman, 1997). Both U.S. and Japanese electronics firms developed their cross-national production networks in Asian countries in 1980s and 1990s. U.S. firms' "open, fast, flexible, formal, and disposable" cross-national production networks have been claimed to defeat Japanese firms' "closed, cautious, centralized, long-term and stable" networks, which helped the U.S. to regain its leadership in the global electronics industry in the 1990s after losing the battle to Japanese firms in the 1980s (Borrus, 1997).

Focusing on the telecom manufacturing industry in China, this paper contends that the existing literature needs to be expanded in order to explain the Chinese case. First, product cycle theory could be applied to explain multinational corporations' strategies of importing and localizing their products in China in order to take advantage of lower labor costs and often more significantly to break barriers to the Chinese market. However, rapid technology changes have made the last step of the product cycle less relevant, i.e. for local subsidiaries and joint ventures of multinational corporations to produce the entire products based on their lower labor costs and technological maturity. These locally made products are then exported to advanced countries, while multinational corporations abandon the production in advanced nations. When the local manufacturers, largely pushed by government regulations and industrial policy, are ready to localize the entire production, Western countries have moved to newer and more advanced products, which makes export opportunities to the Western world vanish. Meanwhile, a new product cycle with a new and more advanced product starts and repeats the previous process when multinational corporations begin to export and localize new products in China. This process works as a "dynamic adding-and-dropping" product cycle (Tan, 2001). Local subsidiaries and joint ventures of multinational corporations have mostly served as low labor cost assembly facilities and a conduit to break entry barriers to local Chinese markets.

Second, there are no significant indicators pointing to local multinational subsidiaries and indigenous manufacturers serving as a substantial part of the cross-national production networks in the global telecom industry yet, although there are some signs of potential development. Most local facilities have served as assembly centers for the Chinese market rather than powerful production centers to produce components, subsystems or even systems for other markets outside China. Detailed examination of this aspect is beyond the scope of this paper.

Third, the success of "Wintelism" and the maturity of cross-national production networks in the global market have had significant impacts on China's indigenous industry. Indigenous manufacturers start to take advantage of their strength in the distribution and production value-chain, such as marketing skills, government supports, locally oriented operating and applications hardware and software design, and low cost assembly. Meanwhile, indigenous manufacturers outsource their weaknesses, such as manufacturing of IC chips and other components and possibly ASIC design to Western corporations. The final products are assembled and marketed by indigenous producers. These products have emerged as strong competitors to multinational corporations both in the Chinese market and in some less-developed markets. This model of "reversed cross-national production networks", firms in developing countries contract out to firms in advanced countries, represents a feasible industrialization path with great potential to enable indigenous manufacturers to emerge as competitors in advanced Western markets as well as less developed markets including China. Findings of this paper are relevant to other large developing countries including India and Brazil where a large domestic market could support the emergence and growth of indigenous manufacturers.

## 2. The Chinese Market and Government Policy

**2.1. Significant Market Size and Impressive Growth Rate**

China has emerged as a significant telecom market in the world since the late 1980s. On average, more than 10 million telephone lines have been added to China's telecom network each year in the last decade. Total telephone lines in China reached to 145 millions in 2000. The wireless phone network in China has experienced an even higher growth rate. In July 2001, China's Ministry of Information Industry (MII) announced that its mobile phone users reached 120.6 million, which makes China the largest mobile communication market in the world, surpassing the 120.1 million users in the United States. A late starter on the Internet, China is trying to catch up, with a 200-300% annual growth rate. China's total Internet users reached 26 million in June 2001. These impressive growth rates are illustrated in Figure 2-1. More significantly, continuous growth across all three fronts has been forecasted even though the rates might slow down.

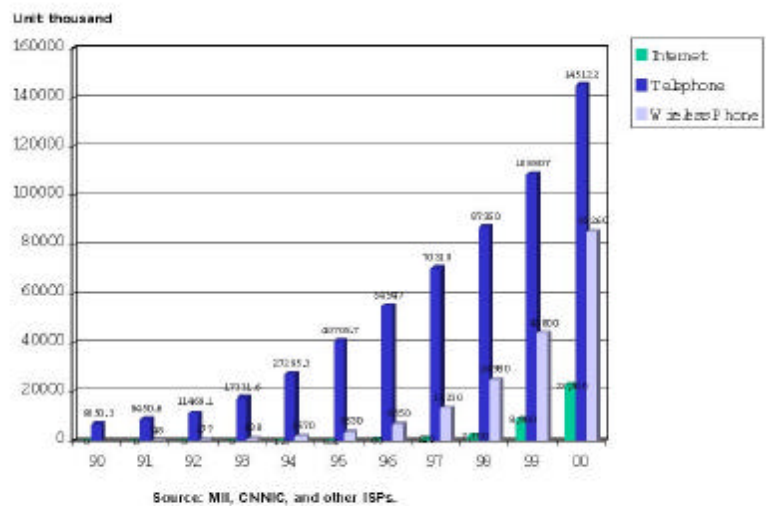

Figure 2-1: Internet, Telephone & Wireless Users in China: 1990-2000

Behind the impressive growth of telephone, wireless phone, and Internet users, there has been a rapid and large-scale facility build-up to support the services. This infrastructure development has created a huge demand for various systems that for most part could not be supplied by indigenous manufacturers. Almost all the significant telecom multinational corporations in the world have been attracted to the Chinese market and have engaged in various activities from direct import, local production, and joint local production to technology transfer. These firms include AT&T (Lucent Technologies), Motorola, CISCO, Nortel, Ericsson, Nokia, Alcatel, Siemens, NEC, Fujitsu, and Samsung among others.

**2.2 Government Policy**

Government regulations and policy have, to a large extent, helped to shape competition in China's telecom manufacturing and service market. China has recognized that involving multinational corporations is part of the solution to national economic and social development (UNTCAD, 1999). But China has apparently also realized that "Development of the host countries is a fortuitous side effect at best, which will only come about if the host government maintains enough autonomy and control to guarantee that the benefits of foreign direct investment are shared between providers and recipients of foreign capital" (Stallings, 1990). Regulations and policy have been actively formulated to promote the production localization of multinational corporations' subsidiaries and joint ventures and to nurture indigenous manufacturers in China's telecom industry.

Relevant policies have focused on foreign trade and investment, product procurement, and favorable financial and R&D support. Foreign trade and investment regulations and policy serve to promote or restrict direct imports and the establishment of local subsidiaries and joint ventures by multinational corporations. Localization of components, subsystems and even entire systems is often specified in these regulations and policies. The most common practice is reflected in the publication of the *'Government Guidelines for Foreign Investment in Telecommunications'* that are updated regularly. The guidelines often divide foreign entries into three categories: encouraged, restricted, and prohibited. The encouraged category consists of products in which China's local capacity lags behind the global market. Direct import is the dominant means to meet market demand. The restricted category includes production that could be supplied by local subsidiaries and joint ventures of multinational corporations and some indigenous manufacturers. Restricting further foreign investment in these products releases the competitive pressure on the exiting local firms and leaves space for them to grow. The prohibited category includes products that use outdated technologies and that target an already saturated market.

Procurement policy has been constantly and effectively deployed in China, especially when there are a small number of larger buyers. Procurement has been continuously used to help acquire technology by giving up market shares to multinational corporations that are willing to transfer technologies and to establish local production. Procurement has also been deployed to promote products of emerging indigenous manufacturers. In addition, policies of providing favorable bank loans, attractive tax breaks, and generous R&D support have also been utilized to promote the localization of production and to nurture the growth of indigenous manufacturers. The large size of China's domestic market puts China in a favorable position to implement these regulations and policies.

Government regulations and policy in advanced countries also significantly influence the market competition in China's telecom market, most notably the policy of blocking direct investments and blocking certain technology exports to China. Taking the U.S. as example, the long and comprehensive list of restricted technologies that could not be exported to China in the 1970s -1980s has had profound impacts on China's IT and telecom industry. The policy could have implications in three perspectives: retarding the growth of indigenous production in China; encouraging China's technology reliance on countries other than the U.S., which eventually jeopardizes trade with U.S. firms; and pushing China to expand and rely more on indigenous R&D capacity. Political conflicts today often lead to the implementation of technology export blocking. The blocking policy is one of the potential threats to the "reversed cross-national production networks" discussed in section 5.

## 3. Evolving Market Competition

While most telecom markets in China start from direct import from multinational corporations, different paths have evolved for different products. Key determining factors are the level of market competition, technology maturity, local government policy implementation, market demand by users, the market power of local service providers, and the capacity of indigenous manufacturers. When the market is very competitive among several multinational corporations, the technology is mature, demand is strong, and government industrial policy can be effectively implemented, direct import is rapidly displaced by local assembly/production through sole-owned subsidiaries or joint ventures of multinational corporations. Otherwise, direct import dominates the market for a longer period. Overall, it takes indigenous manufacturers a long time to catch up with every new product launched by multinational corporations, and the technology complexity and the availability of core components play significant roles in determining the length of the catching-up period.

### 3.1. Central Office Switches

Traditional Central Office Switch market is an instance where competition among many multinational corporations is fierce, technology is mature, demand is strong (more than 10 million lines annually) and China's government industrial policy can be effectively implemented (Shen, 1999). Many global suppliers are keen to enter the Chinese market, including Alcatel, Siemens, Ericsson, Nortel, NEC, Lucent Technologies, and Fujitsu. The intense competition puts China in a favorable bargaining position.

The technology of traditional central office switches is based on the principle of stored programming control, which originated in the 1960s and became mature in the late 1980s and early 1990s. These switches have been deployed on a large-scale in Western countries since the 1970s. Scale of economy limits the production of central office switches to a few large corporations. In China, switches were mostly purchased by a single monopoly buyer, then China Telecom. A single monopoly buyer leads to a better organized decision-making process, which makes it more effective for China to implement its industrial policy through procurement and joint venture negotiation. These market features have determined an evolution with four distinct stages for central office switches in China as shown in Table 3-1: 1) 100% direct import; 2) market split between direct import and local assembly by joint ventures; 3) local assembly by joint ventures with some indigenous production; and 4) market split between local assembly by joint ventures and indigenous production.

Table 3-1: Market Shares of Central Office Switches among Three Groups.

|  | 1982 | 1987 | 1992 | 1997 | 2000 |
|---|---|---|---|---|---|
| **Direct Import** | 100% | 89% | 54% | 5% | 0% |
| **Joint Venture** | 0% | 11% | 36% | 63% | 57% |
| **Indigenous Suppliers** | 0% | 0% | 10% | 32% | 43% |

Source: The former MPT & MEI, the MII, suppliers' annual reports, and author's estimate.

This development is very close to a typical product cycle except for the last step. Theoretically, China should start to export these lower cost switches back to Western countries since the production is localized to a large extent. However, technology advances have eliminated the export opportunities. Western markets have moved to more advanced next generation switches including ATM (Asynchronous Transfer Mode) and IP capable switches. These Chinese-made traditional switches have only been exported to some less-developed countries in Asia, Latin America and East Europe.

On the other hand, the Chinese market has recently started to direct import next generation central office switches, ATM and IP capable ones, from advanced countries. The ATM market is currently dominated by direct imports with some very limited local assembly and indigenous production. If the demand keeps up, the market for ATM and IP capable switches is likely to see a repeat of the four steps seen for traditional central office switches.

### 3.2. Wireless Communications Systems

Wireless communications is a market where competition among several multinational corporations is moderately fierce, demand is strong and China's government industrial policy can be effectively implemented with two state-owned buyers, China Mobile and China Unicom. But wireless technology is relatively less mature. Globally speaking, first generation (1G) wireless systems started their large-scale deployment in the early 1980s. 2G started to replace 1G systems in the early 1990s. 3G systems are expected to enter the market between 2003 and 2005.

China's 1G systems have mostly relied on direct import from two suppliers, Motorola from the U.S. and Ericsson from Sweden. When 2G systems entered the Chinese market, demand picked up strongly. Market competition became fiercer among more suppliers, including Motorola, Ericsson, Nokia, Nortel, Siemens, Alcatel and others. The strong demand and fierce competition, together with a government policy to promote local production, quickly turned China's wireless market from direct import into local assembly, as shown in Table 3-2. However, indigenous manufacturers are catching up slowly, mainly because the technology is new and complex. Products have not been exported back to Western markets because Chinese production is mostly assembly and the multinational corporations have not abandoned their production in advanced countries.

Table 3-2: Market Share of China's 2G Wireless Market.

| Year | | 1994 | 1999 | 2000 |
|---|---|---|---|---|
| Infrastructure Equipment (Base Stations & Mobile Switches) | Direct Import | 100% | 31%* | 25%* |
| | Subsidiaries & Joint Ventures | 0 | 66%* | 70%* |
| | Indigenous Producers | 0 | 3% | 5% |
| Terminal Equipment (mostly handsets) | Direct Import | 100% | 5% | 2% |
| | Subsidiaries & Joint Ventures | 0 | 92% | 88% |
| | Indigenous Producers | 0 | 3% | 10% |

Source: Survey by MII's Telecommunications Information Research Institute. *: estimated by author.

### 3.3. Data Communications Systems

Data communications is a relatively new market in China since large-scale deployment of internal networks only started in the late 1990s in Chinese enterprises and government agencies. Technology for data communications is relatively new and dynamic in the global market. Competition in China's data communications market is mostly limited among several North American multinational corporations including Cisco, Nortel, Lucent and 3Com. The large number of Chinese buyers and the widespread ownership structure of these networks have made the implementation of China's government industrial policy difficult and less effective.

As in other markets, the data communications market started with direct import from advanced countries. The smaller number of suppliers, the complexity and dynamics of product technology, and the large number and diverse ownership of Chinese buyers have created much less severe pressures for multinational corporations to localize their production in China, compared with central office switches and wireless communications equipment. This market situation leads to a direct import-dominated market. For the high-end products including high-speed Ethernet switches, high-performance routers, and ATM switches, direct import completely dominates with an over 98% market share. The low-end market sees some competition between direct import and indigenous production, as shown in Table 3-3.

Table 3-3: Market Share of China's Data Communications Market.

| Year | | 1998 | 1999 |
|---|---|---|---|
| High-end (high-speed Ethernet switches, high-performance routers, and ATM switches) | Direct Import | 99% | 98% |
| | Subsidiaries & Joint Ventures | 0 | 0 |
| | Indigenous Producers | 1% | 2% |
| Low-end (Dial-up modem market only) | Direct Import | 73% | 66% |
| | Subsidiaries & Joint Ventures | 0 | 0 |
| | Indigenous Producers | 27% | 34% |

Source: Survey by MII's Telecommunications Information Research Institute.

## 4. Production Localization - Interrupted Product Cycle

Competition in China's markets of central office switches, wireless communications equipment and data communications equipment has largely followed the prediction of product cycle theory. Telecom products are first developed and deployed in advanced countries. As demand emerges in the Chinese market, these products are brought into China through direct import. When local market entry barriers created by government regulations and policy increase, multinational corporations are pressured to establish local assembly to break the barriers. Government regulations and policy tend to push the local assembly first to partial local production and then to total local production. This process is closely tied to market dynamics including competition among multinational corporations and from indigenous manufacturers. Lower local labor costs are attributed to lower assembly costs. However, lower labor costs are rarely appreciated in overall product R&D. Lower labor costs are not significant in the development and production of high-end components and subsystems since China's workers are not yet sophisticated enough to do this work.

Two things are missing in the product cycle in China's telecom market. First, subsidiaries and joint ventures of multinational corporations are mostly compelled by China's government policy rather than motivated by their own business considerations to localize their productions. Local outsourcing and local design are pursued only passively. Most establishments including the two largest ones, Motorola's sole-owned subsidiary in Tianjin and Alcatel's joint venture in Shanghai, are assembling facilities with limited local components and design capability. None of these facilities has emerged as a significant producer of components and subsystems for multinational corporations' global market, and there are only some small-scale exports of components and subsystems from these Chinese facilities to other markets.

Secondly, the product cycle is incomplete. The last step of a completed product cycle should allow China's local facilities to manufacture the product based on China's lower labor costs. These products should be exported to advanced countries while multinational corporations abandon their production of these products in developed countries. In reality, this rarely occurs in China's telecom market. There are occasions where Chinese facilities are used as an assembly site for the global market. For example, Motorola assembles some of its high-end wireless phones in China by supplying all the components and subsystems. Some of these wireless phones are then sold in the U.S. market. Alcatel has occasionally asked its joint venture in Shanghai to assemble some switches and later to sell them to a less developed third country. However, there is no case where a product is designed, developed, and produced in a Chinese facility of subsidiaries and joint ventures of multinational corporations, which is then sold on a large scale to an advanced market.

Production localization by subsidiaries and joint ventures of multinational corporations has mostly ended with an "interrupted product cycle" as shown in Figure 4-1. Products are first imported to the Chinese market, and then pushed to local assembly and to partial or complete local production. When multinational corporations start to abandon production in advanced countries, markets in advanced countries have also abandoned these products. This situation has eliminated the possibility for subsidiaries and joint ventures of multinational corporations to export their products from China to advanced markets. Meanwhile, new and advanced products appear in advanced markets and are imported to the Chinese market. These products then tend to go through this "interrupted product cycle" again. It is possible that changes in government policy, multinational corporations' strategy and the global market environment alter the "interrupted product cycle" in the future.

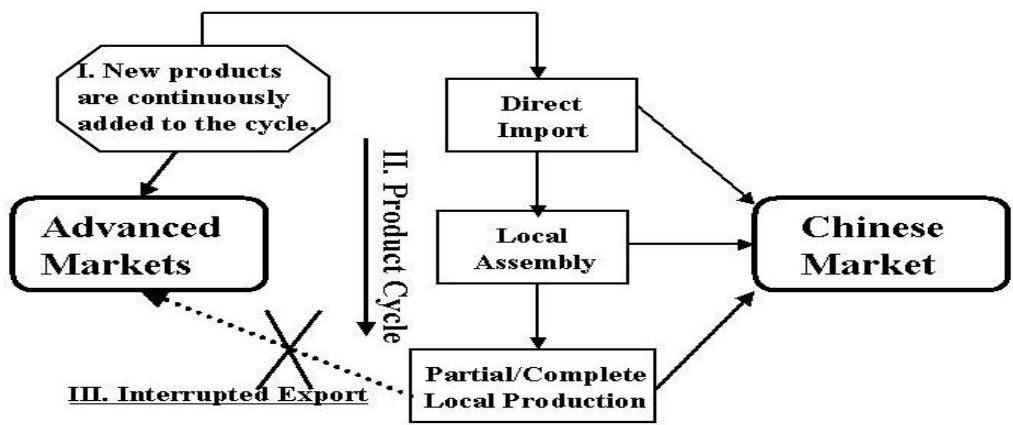

Figure 4-1: The Interrupted Product Cycle.

# 5. Emerging Indigenous Production - Reversed Cross-national Production Networks

While China's telecom markets are mostly dominated by direct import and local assembly by multinational corporations' subsidiaries and joint ventures, market data have demonstrated the emergence of indigenous production. Indigenous production is strong in mature products as well as where local marketing skills are desired and favourable government policies can be implemented. However, 'Wintelism" and Cross-national Production Networks (CPN) have facilitated the growth of indigenous production in countries like China.

**5.1. Elements in the Value-Chain: Dis-integration and Outsourcing**

Wintelism has indicated that traditionally integrated system elements in the production value-chain could be dis-integrated. Elements, including components, subsystems, system assembly, operation software, and applications software, become separate and critical competitive markets. It is argued that Cross-national Production Networks make it possible for the production of any elements in the value-chain to be contracted out to independent producers. These independent producers are located in the global economy (Borrus & Zysman, 1997; Borrus, 1997). Most research has focused on implications of this changing global market environment on multinational corporations in advanced countries. Fundamental influences have affected indigenous industries in developing countries as well.

The emergence of China's indigenous corporations as powerful competitors to multinational corporations, mostly in the Chinese market and increasingly in other developed markets, is partially a result of the success of Wintelism and Cross-national Production Networks. Indigenous corporations have a strong position in almost all the elements of the distribution value-chain. They are close to the local market. They are often protected by government industrial policies including regulations on foreign investments, procurement rules, and favourable financial incentives.

However, indigenous manufacturers are often weak in some elements in the production value-chain. China's fundamental Integrated Circuits (ICs) industry is still far behind the advanced countries. Production of many key components and subsystems, especially high-end ICs could not be conducted locally. China's indigenous manufacturers are, in many cases, not sophisticated enough in some areas of system design and software development.

The separation and dis-integration of elements in the production value-chain allow China's indigenous corporations to contract out their weak areas to U.S., European, and other Asian companies. These contractors are often not the direct competitors of the indigenous corporations in the Chinese market but specialize in different elements of the production value-chain. This relationship leads to a typical distribution of elements in the distribution and production value-chains for emerging Chinese indigenous manufacturers -- keeping their strengths and outsourcing their weaknesses. The distribution is shown in Figure 5-1. This author calls this relationship "reversed cross-national production networks" where firms in developing countries contract out to firms in advanced countries.

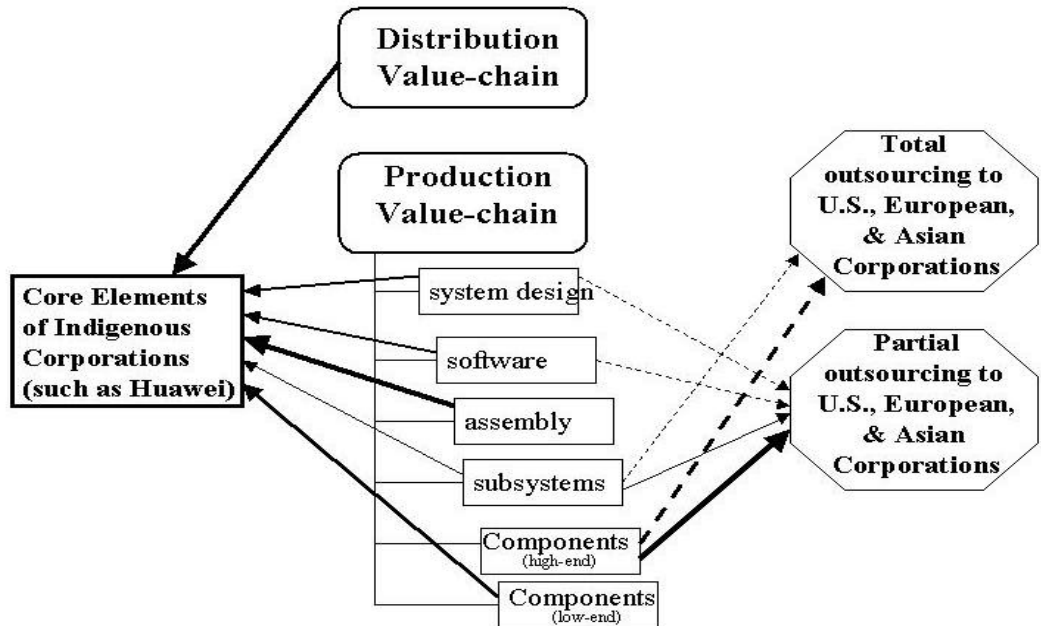

Figure 5-1: A Typical Distribution of Value-chain Elements.

## 5.2. Huawei Technologies as a Case

Major indigenous corporations in China's telecom industry include Huawei Technologies, Zhongxin Corporation, Datang Telephone, and Grand Dragon Telecommunications Corporation. While each of them has been involved in reversed cross-national production networks, Huawei Technologies stands as a typical example. Huawei Technologies (www.huawei.com), established in 1988and headquartered in Shenzhen, China, is a private telecom corporation fully owned by its employees. Total sales of Huawei reached US$2.66 billion in 2000. Its products cover the full range of the telecom industry, including fixed networks, wireless networks, data communications networks and optical transmission networks. It is one of the most successful indigenous Chinese firms in the IT industry. While focusing on the Chinese market, its products have been exported to over 40 countries, including Brazil, Germany, Kenya, Russia, and Thailand.

Many factors have contributed to the miracle of Huawei, including the entrepreneurship of its founders and employees as well as government supports. But one of the significant factors is the changing global market environment because of Wintelism and Cross-national Production Networks. As an indigenous firm, Huawei has many advantages over multinational corporations in the distribution value-chain. On the production side, Huawei is capable of designing and assembling its own systems at a lower cost. However, neither Huawei nor any other Chinese producers are capable of producing Huawei's own designed ASICs (Application Specific ICs). According to Huawei's new release, its ASICs are mostly contracted out to partners among Texas Instruments, IBM, Motorola, Lucent Technologies, Intel, Sun Microsystems and others. These partnerships are mostly in the semiconductor sector, part of the component market where Huawei does not compete. The outsourcing with Motorola and Lucent Technologies is also mostly with their semiconductor divisions.

This kind of cooperation with foreign partners enables Huawei to focus on its own strengths and compensate for its weaknesses through outsourcing. Final products, assembled in Huawei's facilities in China, are telecom systems with up-to-date components and software for the right market, which has led to the great success and rapid growth of Huawei in both the Chinese and global markets. Without such reversed Cross-national Production Networks, Huawei and other indigenous firms could not produce their up-to-date telecom systems.

## 6. Conclusion

As one of the most significant telecom markets in the world, China has attracted almost all the multinational corporations to engage in trade and production. China's telecom market features the coexistence of direct import, local subsidiaries and joint ventures of multinational corporations, and indigenous manufacturers. The interaction among multinational corporations' strategy, government policy and regulations, and indigenous manufacturers' practice has shaped market competition.

In most cases, the "interrupted product cycle" best explains market competition. Telecom products are first developed and deployed in advanced countries and then imported to the Chinese market as demand emerges. Multinational corporations are then pressured to establish local assembly and production facilities to break the market entry barriers created by government regulations and policy. The lower labor costs in China may also play a role in production localization. Localization can be pushed from local assembly to partial or complete local production, since subsidiaries and joint ventures of multinational corporations are rarely motivated by their own business considerations to localize their production. More significantly, the product cycle is often incomplete. When local manufacturers in China are ready to export their products and multinational corporations start to abandon production in advanced countries, markets in advanced countries have also abandoned these products. New and advanced products appear in advanced markets and are imported to the Chinese market. These products then tend to go through this "interrupted product cycle" again.

In contrast to this norm, some indigenous Chinese firms have emerged as powerful competitors mostly in the Chinese market. Their successes reside in the model of "reversed Cross-national Production Networks". Such indigenous corporations have kept their strengths in almost all the elements of the distribution value-chain and some elements of the production value-chain including lower assembly labor costs. Meanwhile, the separation and dis-integration of elements in the production value-chain allow China's indigenous corporations to contract out what they are weak in, especially in IC production, to U.S., European, and other Asian companies. The final products, assembled in China, are often strong competitors to other multinational corporations in the Chinese market and some other less-developed markets.

While the findings of this research are specific for the Chinese market, they are relevant to some other large developing countries such as India and Brazil.